 \documentclass[twocolumn]{svjour3}          
\smartqed  

\usepackage{graphicx}
\usepackage{bm}
\usepackage{epstopdf} 
\usepackage{amssymb}

\usepackage{amsmath}

\usepackage[switch]{lineno}

\usepackage{paralist}
\usepackage{titlesec}

\usepackage{wasysym}
\usepackage{tikzsymbols}
\usepackage{tikz}
\usetikzlibrary{shapes}

\usepackage{xcolor}
\usepackage{cancel}
\usepackage[normalem]{ulem} 
\usepackage{lipsum}

\usepackage{natbib}
\setcitestyle{authoryear,round}
\setcitestyle{citesep={;}}

\usepackage[labelformat=simple]{subcaption}

\usepackage{bbding}

\usepackage{url}
\usepackage[hidelinks]{hyperref}

\graphicspath{{figs/}}

\definecolor{blue1}{rgb}{0.0863, 0.2751, 0.5290}
\definecolor{blue2}{rgb}{0.1662, 0.4838, 0.7258}
\definecolor{blue3}{rgb}{0.4434, 0.6963, 0.8272}
\definecolor{blue4}{rgb}{0.7565, 0.8660, 0.9221}
\definecolor{blue5}{rgb}{0.9506, 0.9668, 0.9769}
\definecolor{blue6}{rgb}{0.9927, 0.8716, 0.7997}
\definecolor{blue7}{rgb}{0.9537, 0.6377, 0.5006}
\definecolor{blue8}{rgb}{0.8210, 0.3360, 0.2774}
\definecolor{blue9}{rgb}{0.6431, 0, 0.1479}
\definecolor{red1}{rgb}{0.5407, 0, 0.1288}
\definecolor{red2}{rgb}{0.7744, 0.2348, 0.2229}
\definecolor{red3}{rgb}{0.9145, 0.5399, 0.4153}
\definecolor{red4}{rgb}{0.9890, 0.8038, 0.7000}
\definecolor{red5}{rgb}{0.9810, 0.9621, 0.9502}
\definecolor{red6}{rgb}{0.8342, 0.9061, 0.9463}
\definecolor{red7}{rgb}{0.5613, 0.7663, 0.8670}
\definecolor{red8}{rgb}{0.2277, 0.5488, 0.7528}
\definecolor{red9}{rgb}{0.1166, 0.3515, 0.6286}

\newcommand\thefont{\expandafter\string\the\font}

\usepackage{lineno}

\newcommand{\RVA}[1]{\textcolor{black}{#1}}

\newcommand{\dc}[1]{\textcolor{black}{#1}}

\usepackage{upgreek}
\usepackage{amsmath}
\usepackage{ulem}

\graphicspath{{./}{./Figure/}}

\journalname{Experiment in Fluids}
\begin{document}

\title{Measuring Mean Spanwise Flow with a Rotating Single Hot-Wire Probe over a Wide Yawed Riblet-Textured Surface}


\author{Y. Xia, S. Parajuli, H. Aliffrananda, D. Chung,
N. Hutchins}

\institute{Y. Xia (\Envelope) \and S. Parajuli \and H. Aliffrananda   \and D. Chung \and N. Hutchins   
\at
Department of Mechanical Engineering, \\University of Melbourne, Parkville, VIC 3010, Australia \\
\email{yu.xia4@unimelb.edu.au}\\
\vspace{0.3mm}\\         
}

\date{Received: date / Accepted: date}

\maketitle

\begin{abstract}
{This study investigates the spatial development and recovery of near-wall flow steering induced by yawed widely-spaced riblets. To accurately measure the mean spanwise velocity in the near-wall region, we developed a unique, customised rotating probe equipped with a single hot-wire. This novel technique demonstrates high sensitivity, reliably detecting small spanwise velocity components. We applied this probe to map the flow over straight-to-yawed and yawed-to-straight riblet configurations. For the straight-to-yawed transition, the downstream distribution of the mean spanwise velocity closely aligns with the \dc{step-forced} spatial Stokes layer (SSL) solution, achieving an equivalent active wall motion of $V^+_\mathrm{equ} \approx 2.0$. This offers a simplified analytical pathway for future passive flow manipulation studies.  Conversely, flow recovery over the yawed-to-straight configuration diverges from the SSL step-down phase prediction. Instead, a spatial lag was observed where crest-adjacent flow recovered quickly, but significant residual spanwise flow ($V^+ > 0.5$) persisted far downstream ($\hat{x}^+ > 1000$).
}
\end{abstract}


\section{Introduction}
\label{intro}

In canonical turbulent boundary layers (TBLs), the mean spanwise velocity component is nominally zero and is therefore often overlooked. However, recent advancements in drag reduction have brought transverse flow dynamics into sharp focus. Near-wall transverse forcing techniques -- such as in-plane wall oscillations or travelling waves -- have been extensively investigated as effective active mechanisms for reducing drag \citep{quadrio_drag_2011,leschziner_frictiondrag_2020a,ricco_review_2021a,ding_acceleration_2024a}. Notably, parametric performance maps \citep{quadrio_streamwisetravelling_2009} reveal that drag reduction efficacy is highly correlated with the near-wall spanwise velocity profile generated by the forcing, \RVA{and more specifically, the rate at which this transverse shear changes, identified as the wall acceleration \citep{ding_acceleration_2024a}.}
Beyond active active control, passively manipulating the spanwise steering flow in the vicinity of the wall, such as deploying surface textures with sinusoidal or wavy riblets, has also demonstrated significant drag reduction benefits \citep{peet_theoretical_2009,kramer_wavy_2010,sasamori_experimental_2014,sasamori_parametric_2017,cafiero_manipulation_2024}. 

Because both active and passive transverse flow control mechanisms are fundamentally governed by near-wall dynamics, obtaining highly accurate spanwise velocity measurements in the extreme near-wall region (e.g. $z^+ < 15$) is a strict prerequisite for validating theoretical models. However, accurately resolving the spanwise flow in the immediate vicinity of the wall presents a formidable experimental challenge. Optical techniques like Particle Image Velocimetry (PIV), while valuable for large-scale flow visualisations, suffer from fundamental limitations near the wall. PIV measurements are frequently plagued by catastrophic laser flare and surface reflections. Although techniques such as fluorescent PIV utilising Rhodamine B can largely mitigate optical blinding \citep[e.g. ][]{aburowin_piv_2025}, PIV remains inherently constrained by the spatial volume averaging of its cross-correlation algorithms. For typical high-Reynolds-number experiments, the minimum interrogation window size acts as a severe spatial low-pass filter, obscuring the critical near-wall spanwise flow features.

To circumvent the limitations of PIV, hot-wire anemometry remains the most viable approach for resolving \RVA{the broad frequency content of the} near-wall velocities. To obtain the spanwise velocity component, it is a natural choice to first consider multi-sensor probes, such as conventional cross-wires (X-wires), which can simultaneously resolve instantaneous streamwise and spanwise velocities. However, deploying these multi-wire probes within thin turbulent boundary layers introduces severe spatial resolution penalties. Specifically, standard X-wires (with the wire plane parallel to the wall) suffer from an unavoidable physical gap between the two sensors in the wall-normal ($z$) direction \citep{baidya_spatial_2019}. While employing a V-shaped cross-wire configuration (where both wires are located at the same $z$-height) can successfully extract spanwise components without a wall-normal offset \citep{philip_spatial_2013}, its larger physical footprint in the wall-parallel ($x$--$y$) plane inevitably compromises spatial resolution. 

Since transverse flow control mechanisms require accurate velocity profiling in an extremely confined near-wall region, the geometric compromises inherent to multi-wire probes render them unsuitable. Therefore, utilising a single wire becomes significantly more advantageous. In fact, the fundamental ability of multi-wire probes to resolve orthogonal velocity components originates entirely from the angular response of a single yawed hot-wire filament \citep{webster_note_1962,champagne_turbulence_1967}. Building on this underlying principle, \citet{fujita_measurement_1968} proposed using a single rotated inclined hot-wire to measure spanwise velocity and Reynolds stress components. By calibrating the sensor's angular response, this technique has been successfully adapted to characterise flow fields in various configurations, such as grid turbulence \citep{sirivat_measurement_1989} and rotor blade wakes \citep{jelly_phaseaveraged_2017}. The primary advantage of using a single wire is its negligible physical size in the wall-normal direction. With only one sensing filament, it provides exceptional spatial resolution, thereby allowing for highly dense measurement points along the $z$-axis. \RVA{Yet, the application of these single-wire methodologies for detecting near-wall spanwise velocity in turbulent boundary layers remains largely unexplored.}

Inspired by the compelling spatial resolution advantages of the single-wire technique, the present study aims to measure the mean spanwise velocity components over wide-yawed riblet textures. We first establish the mathematical methodology for extracting the mean flow angle and spanwise velocities (Section \ref{sec:flowAngleDetectionTechsqr}). The physical implementation of this technique is detailed in experimental setup (Section \ref{sec:wideyawedriblettexture}), which crucially features a novel Z-shaped bracket. Differing from conventional inclined probes that rotate along their longitudinal stem axis, this custom bracket allows a standard normal hot-wire to rotate exclusively around the wall-normal direction ($z$-axis), ensuring proximity to the wall without interference. Following a rigorous validation of this customised measurement approach (Section \ref{sec:validation}), we characterise the detailed spanwise flow features over straight-to-yawed and yawed-to-straight riblet configurations (Section \ref{sec:results}). Based on these high-resolution measurements, we also introduce a novel theoretical perspective: applying the spatial Stokes layer (SSL) solution -- a framework traditionally associated with active wall motion -- to model this passive flow control scheme. Finally the implications of these findings are concluded in Section \ref{sec:discussionconclusion}.

\section{Flow angle detection technique}
\label{sec:flowAngleDetectionTechsqr}

\begin{figure}[t!]
    \centering
   \qquad\quad \includegraphics{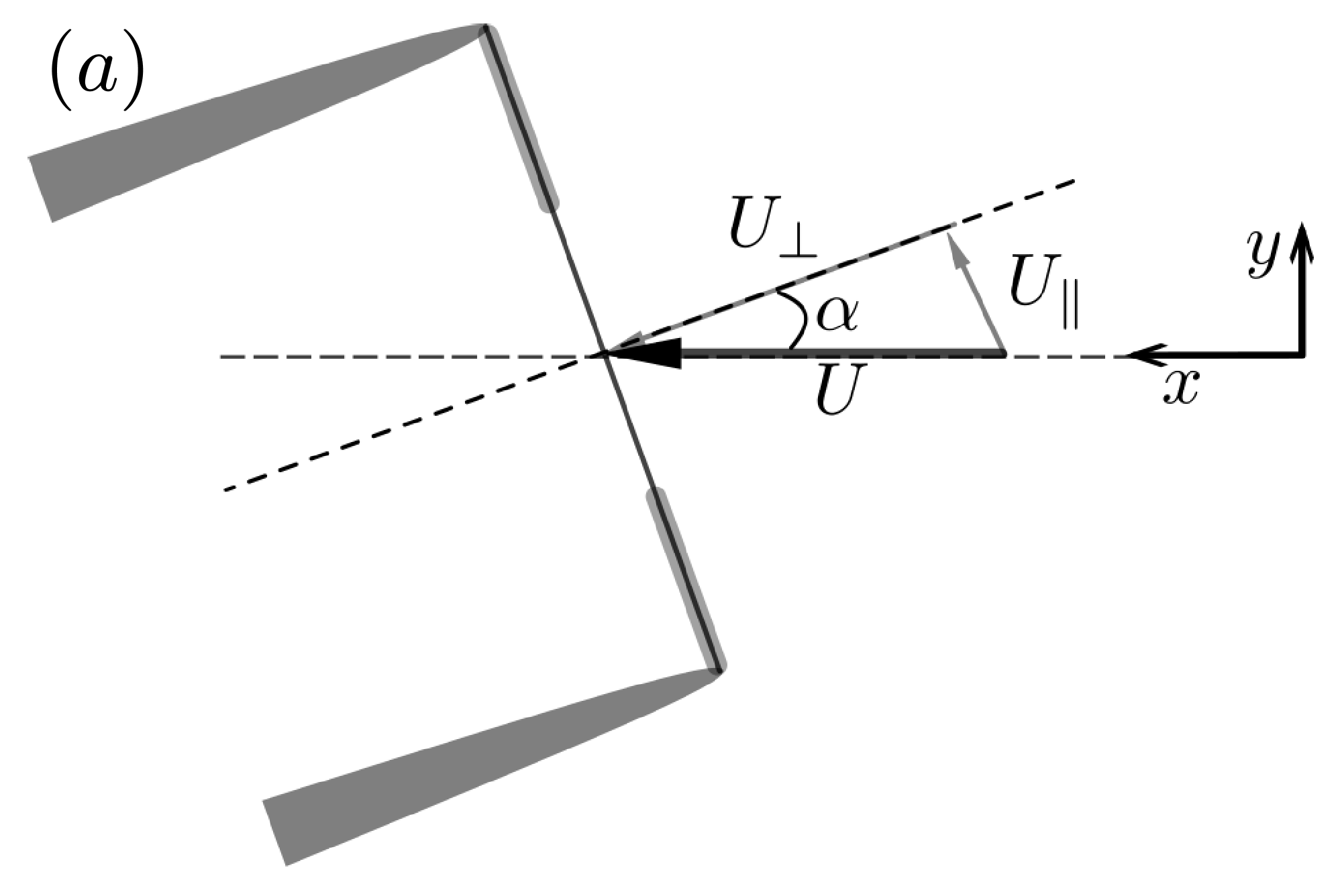}
\\   \includegraphics{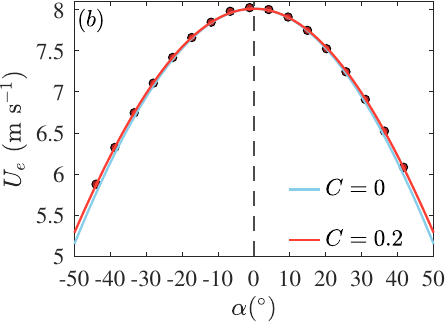}
    \caption{($a$) Schematic of a hot-wire at a yaw angle $\alpha$ within a purely streamwise direction. ($b$) Comparison between  \textit{cosine law} (Eq. \ref{eq:cosineLaw}) and Champagne's relationship (Eq. \ref{eq:champagne}) for a hot-wire deployed in the freestream.}
    \label{fig:champagnRelationship}
\end{figure}

Figure \ref{fig:champagnRelationship}($a$) shows a hot-wire in the $x$--$y$ plane with a yaw angle ($\alpha$) along the streamwise direction ($x$), with rotation about the spanwise axis $y$. In a steady flow environment, the classical \emph{cosine law} assumes that only the wire-normal mean velocity component cools the sensor:
\begin{equation}
    \label{eq:cosineLaw}
    U_e = U_{\perp} = U\cos(\alpha),
\end{equation}
where $U_e$ is the effective mean cooling velocity. 
This relation assumes the flow passes over an infinitely long wire, \RVA{where the tangential component contributes no convective cooling. 
However, for finite-length wires, the cosine law underpredicts the heat loss. \citet{champagne_turbulence_1967a}  demonstrated that end-conduction to the prongs creates a non-uniform temperature distribution along the length of the wire. Because of this longitudinal temperature gradient, the tangential velocity component actively contributes to convective heat transfer, resulting in an excess heat loss that the classical \emph{cosine law} fails to capture.}
To account for this tangential cooling, 
\citet{hinze_turbulence_1975,webster_note_1962,champagne_turbulence_1967a} introduced an empirical yaw factor $C$ into the response equation:
\begin{equation}
U_e^2(\alpha) = U^2_{\perp} + C^2U^2_{\parallel} = U^2\cos^2\alpha+C^2U^2\sin^2\alpha.
\label{eq:champagne}
\end{equation}
The factor $C$ depends primarily on the wire's length-to-diameter aspect ratio ($l/d$), the wire Reynolds number ($Re_d$), the selected overheat ratio, and the probe geometry \citep[][Chapter 2]{hinze_turbulence_1975}. \citet{champagne_turbulence_1967a} measured platinum wires without stubs (unplated) and
\RVA{reported $C \approx 0.2$ at $l/d =200$, with the value decreasing toward zero as $l/d$ approaches $600$, recovering the classical cosine law in the infinite-length limit.}
\RVA{Eq.~(\ref{eq:champagne}) was originally established for steady flow in a low-turbulence jet \citep{champagne_turbulence_1967a}. Its application to turbulent flow rests on a quasi-steady assumption: the thermal response time of the wire is sufficiently short compared with the timescales of the turbulence of interest. Consequently, the steady-state relation is taken to hold at every instant, mapping the fluctuating velocity components directly to the instantaneous effective velocity $u_e(t)$.}
Figure \ref{fig:champagnRelationship}($b$) presents the result of the mean effective velocity of the hot-wire in the freestream as a function of yaw angle ($\alpha$). Here, $U_e = U_\infty$ when the yaw angle is zero. The \emph{cosine law} ($C = 0$) accurately captures the sensor response for $|\alpha| < 30^\circ$, \RVA{but slightly diverges by $2\%$ at $|\alpha| = 45^\circ$, whereas Champagne's relationship (Eq. \ref{eq:champagne} with $C \approx 0.2$) collapses the data across the full angular sweep.}

\begin{figure}[t!]
    \centering
    \qquad\quad \includegraphics{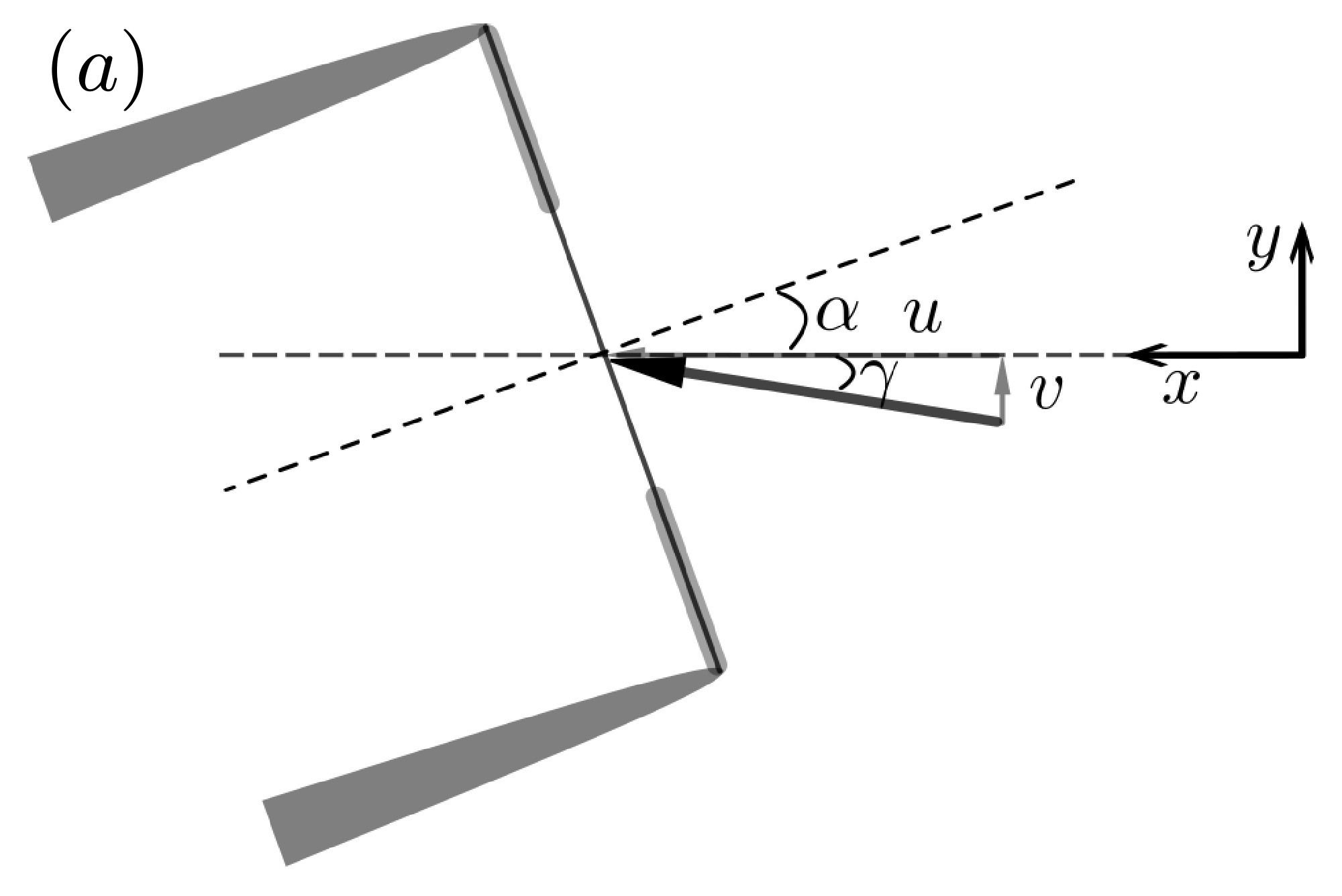}\\
    \includegraphics{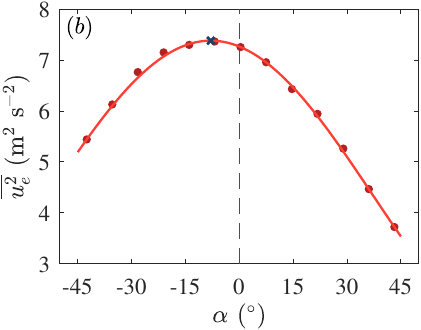}
    \caption{($a$) Schematic of a hot-wire at a yaw angle $\alpha$ in a flow containing a spanwise velocity component. \RVA{$\gamma$ ($=\arctan(v/u)$) is the instantaneous flow angle.} ($b$) Demonstration of the relationship described by Eq. (\ref{eq:ChampagneEalpha}) for a yawed hot-wire subjected to a non-zero spanwise flow. $\times$ marks the peak yaw position.}
    \label{fig:Champagne2}
\end{figure}

Eq. (\ref{eq:champagne}) assumes a flow dominated by the streamwise velocity with negligible spanwise velocity (figure~\ref{fig:champagnRelationship}\textit{a}). \RVA{When a genuine spanwise component ($v$) is present in the $x$--$y$ plane (as shown in figure \ref{fig:Champagne2}$a$), the wire-normal and wire-tangential components must be reconstructed from the instantaneous ($u,v$) via rotation through $\alpha$, giving}
\begin{equation}
u_e^2=(u\cos\alpha-v\sin\alpha)^2+C^2(u\sin\alpha+v\cos\alpha)^2.
\label{eq:champagne2}
\end{equation}
Time-averaging Eq. (\ref{eq:champagne2}) and grouping terms gives 
\begin{equation}
    E(\alpha) \equiv \overline{u_e^2}(\alpha)
    = E_0 + E_c \cos 2\alpha + E_s \sin 2\alpha,
    \label{eq:ChampagneEalpha}
\end{equation}
with coefficients
\begin{equation}
\label{eq:coefficientE}
    \begin{aligned}
        E_0 &= \tfrac12(1+C^2)\left(\overline{u^2}+\overline{v^2}\right), \\
        E_c &= \tfrac12(1-C^2)\left(\overline{u^2}-\overline{v^2}\right), \\
        E_s &= -(1-C^2)\,\overline{uv}.
    \end{aligned}
\end{equation}
\RVA{
The yaw factor $C$ is determined once per probe geometry from a freestream calibration: the probe is rotated through  $\alpha \in [-45^\circ, +45^\circ]$ in increments of ${\Delta \alpha = 5.4^\circ}$, and $C$ is obtained as the value that minimises the residual of Eq.~(\ref{eq:champagne}) fitted to the time-averaged $\overline{u_e}(\alpha)$
data of figure~\ref{fig:champagnRelationship}($b$). At each measurement station in the boundary layer, the constant-temperature anemometer (CTA) voltage is converted to instantaneous effective velocity $u_e(t)$ using a fourth-order polynomial fit obtained from a separate velocity calibration at $\alpha = 0$ in the freestream.}
The probe is then rotated through a yaw sweep of $n$ angles $\alpha_i$ ($i = 1,\dots,n$), and at each angle $E(\alpha_i) = \overline{u_e^2}(\alpha_i)$ is computed from time-averaging. The three coefficients $(E_0, E_c, E_s)$ are then estimated by linear least-squares regression of $E(\alpha_i)$ against the regressors $(1, \cos2\alpha_i, \sin2\alpha_i)$.

 Figure \ref{fig:Champagne2}($b$) illustrates \RVA{an example fit at a near-wall location above wide yawed riblets (Section \ref{sec:validation})}, where the riblets induce a substantial mean spanwise flow.
 \RVA{The fitted curve (solid red) captures the data well, and the peak of $E(\alpha)$  is offset from zero (marked $\times$).}
\RVA{
Once the fit coefficients are dtermined, Eq.~(\ref{eq:coefficientE}) can be inverted to yield the second moments directly:
\begin{equation}
\label{eq:secondMoments}
    \begin{aligned}
        \overline{u^2} &= \frac{E_0}{1+C^2} + \frac{E_c}{1-C^2}, \\
        \overline{v^2} &= \frac{E_0}{1+C^2} - \frac{E_c}{1-C^2}, \\
        \overline{uv}  &= -\frac{E_s}{1-C^2}.
    \end{aligned}
\end{equation}
To extract the mean components $(\overline{u}, V)$, we apply the Reynolds decomposition:
\begin{equation}
    \overline{u^2} = U^2 + \overline{u'^2}, \quad
    \overline{v^2} = V^2 + \overline{v'^2}, \quad
    \overline{uv} = U\,V + \overline{u'v'}.
    \label{eq:ReynoldsDecom}
\end{equation}
Substituting these into Eq. (\ref{eq:secondMoments}) presents a closure problem, as three turbulent stresses are introduced.
We close the system by assuming the moderate turbulence intensities in the flow field (i.e. the flow over wide yawed riblets), specifically $\overline{u'^2} \ll U^2$ and $\overline{u'v'} \ll U\, V$. The mean components then follow as 
}
%
%
\begin{equation}
\label{eq:mean_vel_components}
\begin{aligned}
    U \approx \sqrt{\overline{u^2}} &= \sqrt{\frac{E_0}{1 + C^2} + \frac{E_c}{1-C^2}}, \\[12pt]
    V \approx \frac{\overline{uv}}{\sqrt{\overline{u^2}}} &= \frac{-E_s\sqrt{1+C^2}}{\sqrt{(1-C^2) [E_0(1-C^2) + E_c(1+C^2)]}}.
\end{aligned}
\end{equation}

\section{Experimental setup}

\label{sec:expsetup}

\begin{figure*}[h!]
    \centering
 \includegraphics{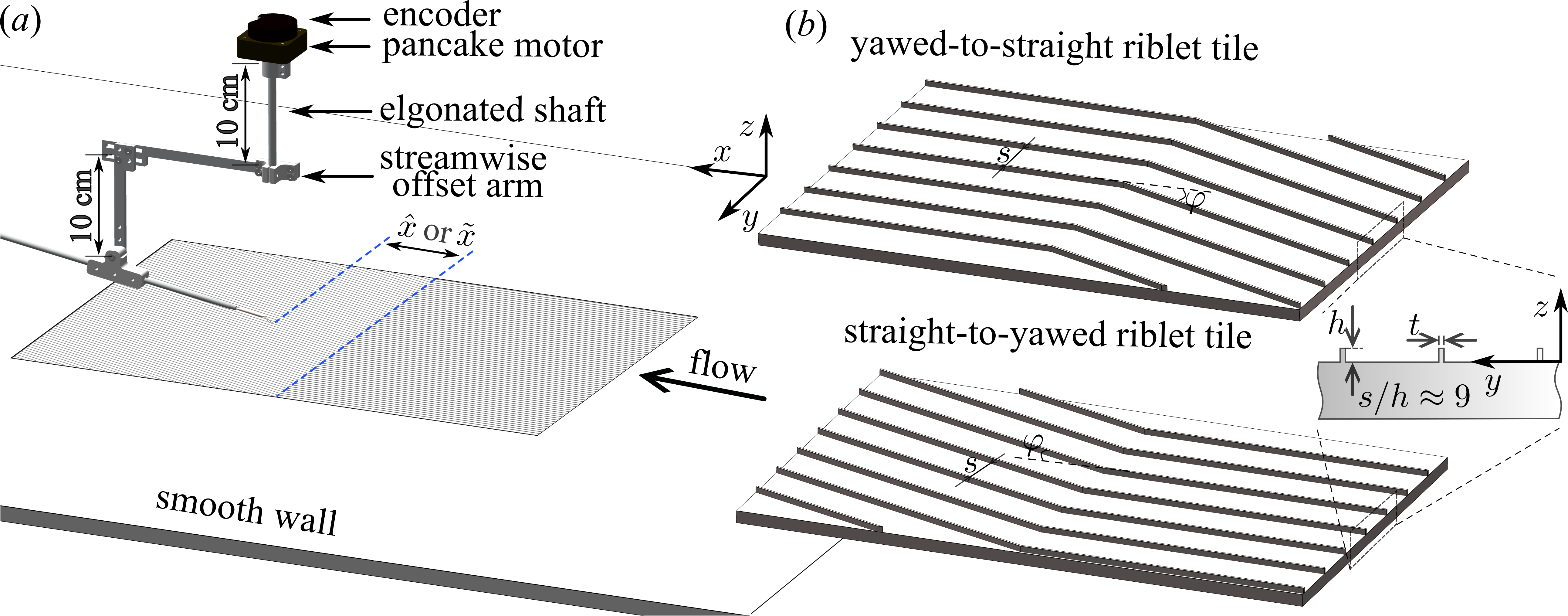}
    \caption{($a$) Schematic of the rotating hot-wire positioned over wide yawed riblets. \RVA{The fetch lengths from the transition point $\hat{x}$ and $\tilde{x}$ denote the straight-to-yaw and yaw-to-straight riblet configurations, respectively.}
    ($b$) The two tiles, featuring the straight-to-yaw and yaw-to-straight configurations, are deployed at $x = 3.75$ m in the wind tunnel. \RVA{For clarity, the riblet elements on both tiles are enlarged by a factor of ten, so that fewer blades are shown than are present on the actual tiles; the schematic is intended to convey the configuration.} Here, $s$ is the riblet spacing and $\varphi$ denotes the yaw angle ($\varphi \approx 20^\circ$). The inset of the $y$--$z$ plane details the cross-sectional geometry of the riblet blade, illustrating the crest height ($h$) and blade thickness ($t$).
    }
    \label{fig:expsetup}
\end{figure*}

\subsection{Rotating probe}
\label{sec:rotatingProbe}
To measure the mean spanwise velocity over yawed riblet tiles, a custom rotating hot-wire probe was developed, as shown in figure \ref{fig:expsetup}($a$).
The system comprises a stepper motor (Sanyo Pancake Stepper Motor SS2421) featuring an embedded encoder (4000 counts per revolution), an elongated shaft, a custom \RVA{streamwise offset arm (dog-leg bracket)} and a Dantec probe support (55H21) holding a single-normal boundary-layer type hot-wire sensor (55P15). The sensor is in-house fabricated from acid etched Wollaston wire to yield a 2.5 $\upmu$m diameter platinum filament with length of approximately 0.5 mm.
The stepper motor rotates the probe about the $z$-axis, providing an angular resolution of 0.09$^\circ$.
To prevent the motor causing local blockage near the wall and consequently corrupting the velocity profiles  within turbulent boundary layers -- a 10 cm elongated shaft offsets the motor so that it is located in the freestream at $z \approx 3.5 \delta$. 

An alternative and mechanically simpler approach would be to use a vertical probe configuration with the stem axis aligned parallel to the $z$-axis (see figure \ref{fig:verticalHotwire}\textit{b} in Appendix \ref{app:vertical} for schematic). However, this orientation places the prongs perpendicular to the mainstream flow, introducing severe aerodynamic interference that artificially accelerates the measured streamwise velocity. Consequently, the conventional wall-parallel orientation was maintained and the streamwise offset arm was deployed. This bracket ensures the hot-wire rotates concentrically about the $z$-axis, aligning the filament centre precisely with the motor's axis of rotation in the $x$- and $y$-axes. A detailed comparative analysis of the aerodynamic interference caused by the vertical configuration is provided in Appendix \ref{app:vertical}.

Following the final assembly, \RVA{the hot-wire was positioned at the mid-point between two adjacent riblet tips, with the wall-normal origin ($z = 0$) set at the riblet valley (i.e. the substrate floor). }
A DSLR camera was positioned facing the probe to quantify any potential spatial drift during rotation ($-45^\circ\lesssim\alpha \lesssim 45^\circ$). To properly assess this drift relative to the surface topography, we define a local coordinate system ($x'$, $y'$) aligned with the yawed riblet blades, where $x'$ is parallel and $y'$ is orthogonal to the blades. These are related to the global streamwise and spanwise directions by $x' = x\cos\varphi - y\sin\varphi$ and $y' = x\sin\varphi + y\cos\varphi$, where $\varphi$ is the riblet yaw angle.  
The maximum spatial drift was restricted to ${\Delta y' \lesssim 0.2}$ mm (i.e.\ ${\Delta {y'}^+ \lesssim 4}$) and $\Delta x'\lesssim 0.05$ mm (i.e.\ $\Delta {x'}^+ \lesssim 1$). Maintaining \RVA{this strict axial alignment} is critical; larger \RVA{radial runout} (e.g.\ $\Delta y' \approx 1$ mm) severely distort the effective velocity response, compromising the least-squares fit of Eq. \eqref{eq:ChampagneEalpha} and yielding erroneous mean spanwise velocities. Appendix \ref{app:Driftrotating} presents an example of measurement corruption due to the rotational drift.


\subsection{Wide-yawed riblet texture}
\label{sec:wideyawedriblettexture}

\begin{table*}[t!]
\centering
\renewcommand{\arraystretch}{1.4}  
\setlength{\tabcolsep}{10pt}        
\begin{tabular}{ccccccccccc}
\hline
 $U_\infty$   & $\nu/U_\tau$ & $s$   & $h$   & $t$   & $s^+$ & $h^+$ & $s/h$ & $t^+$ & $Re_\tau$ & $\varphi$ \\
 (m s$^{-1}$) & ($\mu$m)     & (mm)  & (mm)  & (mm)  &       &       &       &       &           & ($^\circ$) \\
\hline
 8 & 47.6 & 3.96 & 0.44 & 0.15 & 83.2 & 9.2 & 9.0 & 3.2 & 1200 & 20 \\
\hline
\end{tabular}
\caption{Experimental flow parameters and geometric dimensions for the wide-yawed riblet tiles. Here, $U_\infty$ is the freestream velocity, and $\nu/U_\tau$ represents the viscous length scale (wall unit) derived from the baseline smooth-wall measurements. For the riblet geometry, $s$ is the riblet spacing, $h$ denotes the crest-to-valley riblet height, and $t$ is the blade width. The superscript `$+$' indicates normalisation using inner scaling. The spacing-to-height ratio is given by $s/h$, and $\varphi$ represents the yaw angle relative to the streamwise direction.}
\label{tab:yawedRiblet}
\end{table*}

All experiments were conducted in the open-return wind tunnel in the Walter Bassett Aerodynamics Laboratory at the University of Melbourne \citep{xia_highfidelity_2025a}. The tunnel has a working section of 6.5 m $\times$ 0.94 m $\times$ 0.38 m in the streamwise ($x$), spanwise ($y$) and wall-normal ($z$) directions. Throughout the experimental runs, the freestream velocity was maintained at $8$ m s$^{-1}$.

We used a CNC machine to manufacture the blade riblets from acetal, configured into tiles measuring 430 mm $\times$ 290 mm. As illustrated in figure \ref{fig:expsetup}($b$), these riblets feature a cross-sectional spacing-to-height ratio of 
$s/h \approx 9$, differing from conventional riblets (which typically have $s/h \approx 2$). Here, $s$ denotes the riblet spacing and $h$ represents the riblet height from valley to crest. The yaw angle ($\varphi$) of the riblets was set to $20^\circ$ relative to the freestream flow $x$-direction. This specific angle was chosen as it \RVA{generates} strong spanwise flow \citep{nurwahyu2025generating}. \citet{nurwahyu2025generating} also demonstrated that narrower blade widths  ($t$, e.g.\ $t^+ = 1.6$) yield stronger mean spanwise velocities. However, reducing $t$ makes the riblet blades increasingly fragile, posing significant manufacturing challenges. To balance structural integrity with effective flow steering, we selected a blade width of $t \approx 0.16 $ mm ($t^+ \approx 3.2$).

To facilitate the measurements, the manufactured tiles were embedded into the wind tunnel floor at $x = 4$ m and the rotating probe was deployed over the tile, as demonstrated in figure \ref{fig:expsetup}($a$). 
Testing both the straight-to-yawed and the yawed-to-straight tiles allowed us to investigate both the development and the recovery of the flow over the yawed riblets. Table \ref{tab:yawedRiblet} outlines the complete experimental parameters and riblet geometries. It should be noted that the friction velocity ($U_\tau$) and the wall units used for normalisation throughout this study are derived from a baseline smooth-wall case, measured via an oil-film technique \citep[more details in][]{ramani_assessment_2024} at the identical streamwise measurement location.

\subsection{Validation of spanwise flow measurement technique}
\label{sec:validation}
\RVA{To validate the rotating probe, we compare the experimental measurements to a minimal channel direct numerical simulations (DNS) over similar widely-spaced riblets, with $h^+ = 8$, $s/h \approx 9.2$, $t^+ = 1.6$ and $\varphi = 22.3^\circ$ . For comparison, the DNS results and the experimental data will be compared at the midpoint of the riblet valley spanning in the wall-normal direction. To accurately account for the finite sensor resolution, we spatially average the DNS data along a simulated wire length \citep{xia_evaluation_2024}.}
Assuming a hot-wire viscous length of $l^+\approx 11$, the spatially averaged DNS data at the centre of the riblet spacing is processed to calculate the effective velocity based on Eq. (\ref{eq:champagne2}) with $C=0.2$ across various simulated hot-wire yaw angles, $-45^\circ < \alpha < 45^\circ$.

\begin{figure}[!t]
    \centering
    \includegraphics{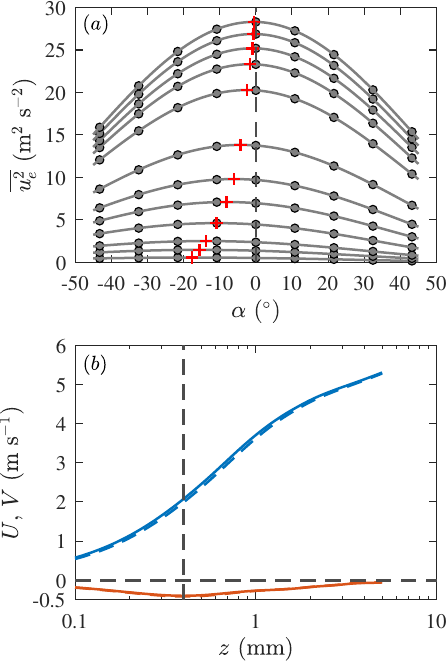}
    \caption{DNS-simulated rotating hot-wire data, with velocities and location converted to physical units. ($a$) Variation of mean-squared effective velocity $\overline{u^2_e}$ with yaw angle $\alpha$ for wall-normal locations $z = 0.1$ to 4 mm.
    The DNS data (calculated via Eq. \ref{eq:champagne2}, denoted by filled circles) are used to produce the least-square fits (solid curves Eq. \ref{eq:ChampagneEalpha}), with the resulting peak yaw positions $\alpha_p$ marked with $+$ at each height. 
    ($b$) Profiles of the streamwise (blue) and spanwise (red) mean velocity components. The original velocity from DNS database (solid lines) is shown alongside the estimations (dashed lines) derived from Eq. (\ref{eq:mean_vel_components}). The vertical dashed line denotes the riblet crest and the horizontal dash line denotes zero velocity at the ordinate. 
    }
    \label{fig:uEffVelocityValidation}
\end{figure}

Figure \ref{fig:uEffVelocityValidation}($a$) presents the \RVA{DNS extracted} mean-squared effective velocity as a function of yaw angle $\alpha$ (ranging from $-45^\circ$ to $45^\circ$) at different wall-normal locations. Note that the dimensionless velocity from the DNS has been converted to physical units according to table \ref{tab:yawedRiblet}.
A least-squares fit to Eq. (\ref{eq:ChampagneEalpha}) is implemented to determine the coefficients $E_0$, $E_c$ and $E_s$. The magnitude of the peak yaw angles ($\alpha_p$), marked with $+$ symbols, decrease as the wall-normal distance $z$ increases, indicating that the flow-steering effect weakens further from the wall. Using the derived coefficients $(E_0, E_c, E_s)$, the mean streamwise  ($U$) and spanwise ($V$) velocities are estimated via Eq. (\ref{eq:mean_vel_components}). Figure \ref{fig:uEffVelocityValidation}($b$)
compares the velocity profiles from the raw DNS data against these estimations. The strong agreement \RVA{within $0.5\%$} validates the rotating hot-wire technique, and subsequent assumptions in deriving Eq. (\ref{eq:mean_vel_components}), for measuring mean spanwise velocity near the wall over the wide yawed riblets. 

\begin{figure}[!t]
    \centering
    \includegraphics{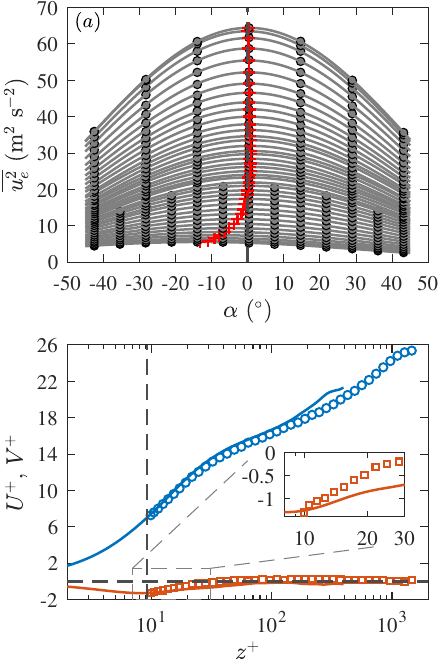}
    \caption{Experimental rotating hot-wire data. ($a$) Variation of mean-squared effective velocity $\overline{u^2_e}$ with yaw angle $\alpha$ for wall-normal locations. Filled circles represent the experimental data at each yaw angle. The solid curves are obtained from linear least-square fits (Eq. \ref{eq:ChampagneEalpha}), with the resulting peak yaw positions $\alpha_p$ marked with $+$ symbols. 
    ($b$) Profiles of the streamwise (blue) and spanwise (red) mean velocity components. The solid lines represent DNS data and the symbols represent measurement results. The horizontal dashed line denotes zero velocity at the ordinate. The vertical dashed line denotes the riblet crest. The inset zooms in profiles in the range of $8 < z^+ < 30$.}
    \label{fig:uEffVelocityEXPValidation}
\end{figure}

To further substantiate this measurement, we compare the DNS results with measured data collected using a rotating hot-wire. The rotating hot-wire was deployed far downstream of the straight-to-yaw transition location, at a fetch length of $\hat{x}^+ \approx 2000$, \RVA{where $\hat{x}$ is distance downstream of the transition point of the riblet tile as defined in figure \ref{fig:expsetup}\textit{a}}. This extensive fetch length was chosen to approximate `fully developed' flow conditions and enabled meaningful comparison with the DNS data that have an effective infinite fetch length \RVA{due to periodic streamwise boundary conditions} ($\hat{x}^+ \rightarrow \infty$).
Figure \ref{fig:uEffVelocityEXPValidation}($a$) plots the experimentally measured $\overline{u^2_e}$ as a function of $\alpha$, suggesting that Eq. (\ref{eq:ChampagneEalpha}) performs well to fit the experimental data. As observed previously from the simulation data, the peak yaw angles (marked by $+$ symbols) approach zero with increasing height, reflecting the diminishing spanwise flow. The least-square fit provides the coefficients ($E_0, E_c, E_s$) necessary to estimate $\overline{u}$ and $V$, as shown in figure \ref{fig:uEffVelocityEXPValidation}($b$) alongside the DNS data (solid lines) and experimental data (symbols). All quantities are non-dimensionalised with viscous scales.
The measured mean streamwise and spanwise velocities collapse well with the DNS data, further verifying the accuracy and utility of the rotating hot-wire method for obtaining flow statistics. 
A minor difference in the mean spanwise velocity is observed near the wall ($8 < z^+< 30$), as highlighted in the inset of figure \ref{fig:uEffVelocityEXPValidation}($b$).
This deviation likely stems from a slight difference in the yaw angle: the DNS features a riblet yaw angle of $22.3 ^\circ$, whereas the experimental setup uses a $\varphi = 20^\circ$ angle. Additionally, the difference in fetch length ($\hat{x}^+ \rightarrow \infty$ for DNS versus $\hat{x}^+ \approx 2000$ for the experiment) plays a role, as the magnitude of the mean spanwise velocity \RVA{and the penetration depth of this spanwise flow away from the surface is inherently smaller}
in a developing flow than in a fully developed one. This is discussed in further detail in Section \ref{sec:results}.

\section{Results}
\label{sec:results}

\begin{figure*}
    \centering  
    \includegraphics{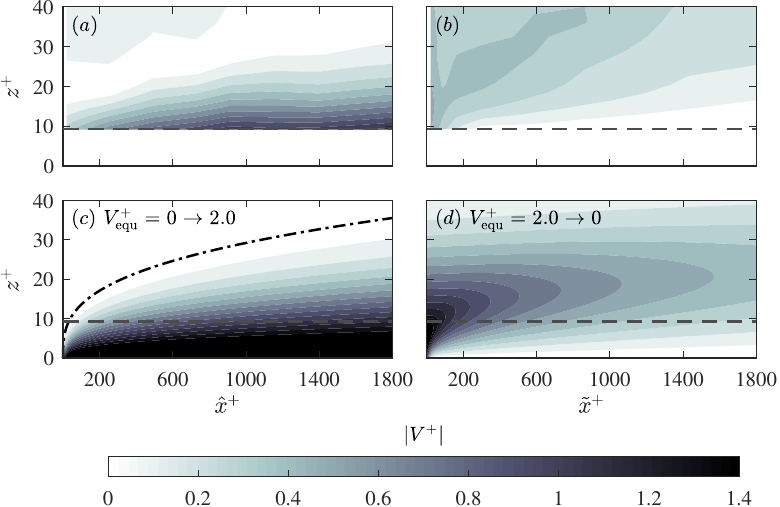}
    \caption{Distributions of the mean spanwise velocity over straight-to-yawed and yawed-to-straight riblet tiles. Measured velocity distributions are shown for ($a$) the straight-to-yawed configuration as a function of $\hat{x}^+$ and $z^+$, and ($b$) the yawed-to-straight configuration as a function of $\tilde{x}^+$ and $z^+$. Panels ($c$) and ($d$) display the corresponding SSL solutions for the straight-to-yawed and yawed-to-straight cases, respectively. The horizontal dashed lines in all panels denote the crest of the manufactured riblet tile. The dash-dotted line in ($c$) denotes the penetration depth $\delta^+_p$.}
    \label{fig:VContourf2x2}
\end{figure*}

We investigated the streamwise development of near-wall spanwise flow
induced by an abrupt change from straight-to-yawed riblets at $\hat{x} = 0$ (see figure \ref{fig:expsetup}\textit{a}).
Wall-normal boundary layer traverses were acquired at multiple streamwise locations, thereby varying the fetch length $\hat{x}$. Figure~\ref{fig:VContourf2x2}(\textit{a}) shows the distribution of mean spanwise velocity magnitude, $|V^+|$, as a function of $z^+$ and $\hat{x}^+$. The induced spanwise flow decays rapidly with increasing $z^+$, confirming that the riblet steering effect is confined to the near-wall region. Immediately downstream of the transition, the footprint is \RVA{confined to the region adjacent to the riblet crest plane. However, increasing $\hat{x}^+$ leads to an intensified near-wall steering, with increased penetration depth further from the wall.
}

In order to describe the wall-normal penetration of flow steering in a simple yet physically rigorous way, we adopt the analytical spatial Stokes layer (SSL) framework proposed by \citet{viotti_streamwise_2009}. While the SSL was originally derived for active control with transverse wall oscillations \citep{quadrio_drag_2011}, we adapt this framework for the present passive scheme through an equivalent boundary condition. Specifically, the passive topographical steering induced by the wide yawed riblet is abstracted as an equivalent active spanwise slip velocity\footnote{The equivalent spanwise slip velocity is adjusted to optimise the model fit to the experimentally measured data is discussed subsequently.}, $V_\mathrm{equ}$. \RVA{Whereas the classical SSL assumes a periodic forcing, the yawed riblet forces the flow only over a localised streamwise region, beginning abruptly at its leading edge. Solving the linearised spanwise momentum equation for this forcing (Appendix~\ref{app:SSL}) gives the
spanwise velocity distribution
\begin{equation}
  V_{\mathrm{step}}^+(\hat{x}^+,z^+) = V_{\mathrm{equ}}^+\left(1 - \frac{1}{\Gamma(4/3)}
  \int_0^{\eta}\exp(-r^3)\,\mathrm{d}r\right),
  \label{eq:maintextSSLsolution}
\end{equation}
where $\eta = {z^+}/{(9\hat{x}^+)^{1/3}}$. Its penetration, set by the height at which $V^+$ has decayed by $99\%$, grows downstream as $\delta_p^+ \approx 2.92\,(\hat{x}^+)^{1/3}$.
}

Figure \ref{fig:VContourf2x2}(\textit{c}) presents this SSL solution (Eq. \ref{eq:maintextSSLsolution}) for an abrupt change in the equivalent wall forcing
from $V_\mathrm{equ}^+ = 0$ to $V_\mathrm{equ}^+ \approx 2$, which reproduces the induced mean
spanwise velocity field downstream of the transition. The horizontal dashed line again
marks the riblet crest. \RVA{The penetration depth is denoted by the dash-dotted line.}
Comparing the measured velocity in figure \ref{fig:VContourf2x2}(\textit{a}) with the SSL solution in figure \ref{fig:VContourf2x2}(\textit{c}), both display the same wall-normal decay and downstream thickening, which supports the use of the SSL framework to describe the passive steering produced by the riblets. Specifically, above the riblet crest, the mean spanwise velocity driven by the equivalent active wall motion closely follows that of the passive straight-to-yawed case. We exploited this correspondence to fix $V_\mathrm{equ}^+$,
selecting the value that minimised the difference between the two fields above the crest;
the closest agreement was obtained at $V_\mathrm{equ}^+ \approx 2.0$.

It should be noted that Eq. \ref{eq:maintextSSLsolution} is rigorously derived for the linear mean velocity profile ($U^+ = z^+$) valid within the viscous sublayer (e.g.\ $z^+ \lesssim 5$). \RVA{For the step change in wall forcing considered here, however, the penetration depth grows with fetch as $\delta_p^+ \approx 2.92\,(\hat{x}^+)^{1/3}$ and soon extends beyond this linear region.
The linearised equation therefore loses strict validity in the buffer layer and above,
where the shear departs from linearity and turbulent eddy viscosity becomes significant.
Even so, it provides a useful qualitative and leading-order quantitative description of
the upward penetration of the flow steering observed in the experiments.}

Following the analysis of flow steering over the straight-to-yawed configuration, we examine the subsequent flow recovery over a yawed-to-straight abrupt change. \RVA{The near-wall spanwise flow downstream of a yawed-to-straight tile
(figure~\ref{fig:expsetup}\textit{b}) is measured at several streamwise stations to
determine the fetch required for the steered flow to realign with the mainstream
direction.}

Figure~\ref{fig:VContourf2x2}(\textit{b}) shows the measured mean spanwise velocity
$V^+$, obtained with the custom rotating probe, as a function of $z^+$
and downstream distance $\tilde{x}^+$ from the transition. The recovery proceeds with a
pronounced wall-normal lag. \RVA{Close to the riblet crest, the fluid responds almost immediately to the removal of the spanwise forcing with $V^+$ quickly trending towards zero.}
The flow further from the wall exhibits a delayed response, retaining noticeable spanwise velocity (e.g.\ $V^+ > 0.5$) well downstream of the transition. Notably, this residual spanwise flow is persistent, remaining measurable beyond at $\tilde{x}^+ \approx 1000$.
\RVA{To complement these measurements, the recovery is modelled with the SSL solution given
in Eq.~(\ref{eq:patchSimpApp}) and shown in figure~\ref{fig:VContourf2x2}(\textit{d}) as a
function of $\tilde{x}^+$ and $z^+$. Here $\tilde{x}^+$ is the distance downstream of an
abrupt change in the equivalent slip velocity from $V^+_\mathrm{equ} \approx 2.0$ to
$V^+_\mathrm{equ} = 0$.}.

\RVA{Although the SSL reproduces the qualitative wall-normal lag of the recovery, the shape of its contours differs markedly from that of the spanwise flow induced by the yawed-to-straight riblets. The spanwise motion driven by the equivalent wall forcing from $V^+_\mathrm{equ}$ to zero decays rapidly with both height and downstream distance. 
}
The SSL model over-predicts the magnitude of $V^+$ throughout the recovery region, whilst the physical passive control generated by the riblets produces a much lower amplitude response.
Ultimately, whereas the SSL effectively describes the mean spanwise velocity distribution for the passive
straight-to-yawed pattern, it fails to accurately capture the recovery of the yawed-to-straight
pattern (figure~\ref{fig:VContourf2x2}\textit{b} and \textit{d}).

\section{Conclusion}
\label{sec:discussionconclusion}
In this study, we developed a unique, customised rotating probe with a single hot-wire to measure the mean spanwise velocity over wide-yawed riblets in the near-wall region. Aided by a DNS database, we validated that this technique is highly sensitive and capable of detecting very small mean spanwise velocity components, particularly at small $z^+$ distances from the wall. We then employed this rotating probe to investigate the development and recovery of near-wall flow steering behaviour over straight-to-yawed and yawed-to-straight riblet configurations, respectively. For the straight-to-yawed riblets, we found that the isolated impulse-forced SSL solution closely resembles the downstream distribution of the measured mean spanwise velocity from the transition point. 
Notably, this represents the first application of the SSL solution to describe a passive control mechanism. Within the SSL framework, an active wall motion of ${V^+_\mathrm{equ} \approx 2.0}$ produces a spanwise steering effect nearly identical to that of the physical passive control, providing crucial theoretical clues and a simple analytical pathway for future studies on passive flow manipulation. 
In contrast to the straight-to-yawed configuration, the flow recovery over the yawed-to-straight riblet tile does not resemble the step-down phase of the isolated impulse-forcing SSL solution from the transition point downstream. The measurement data reveals a spatial lag in the flow recovery: while the region close to the riblet crest responds more rapidly, a residual spanwise flow persists. For example, even far downstream at $\hat{x}^+ > 1000$, the mean spanwise velocity remains significant, with $V^+ > 0.5$.

\section{Acknowledgements}
The authors acknowledge the surface manufacturing contributions at the pilot testing stage from Dr. Will and Mr. Tierney.

\section*{Declaration}
\textbf{Ethical approval} ~Not applicable.
\vspace{.1cm}

\noindent\textbf{Competing interests} ~The authors report no conflict of interest.
\vspace{.1cm}

\noindent\textbf{Authors' contributions} Y.X. analysed the data, prepared the figures and wrote the manuscript. S. P. contributed drafting of Appendix B. N.H. conceptualised the research and developed the key ideas.  H.A. and D.C. provided the DNS database. D.C. and N.H. provided supervision and reviewed and edited the manuscript. All the authors reviewed the manuscript.
\vspace{.1cm}

\noindent\textbf{Funding} The authors acknowledge the support of the Air Force Office of Scientific Research under award number FA2386-23-1-4071.
\vspace{.1cm}

\noindent\textbf{Availability of data and raw materials} Not applicable.


\bibliographystyle{spbasic}      
\bibliography{referenceshort}

@article{baidya_spatial_2019,
  title = {Spatial Averaging Effects on the Streamwise and Wall-Normal Velocity Measurements in a Wall-Bounded Turbulence Using a Cross-Wire Probe},
  author = {Baidya, R and Philip, J and Hutchins, N and Monty, J P and Marusic, I},
  year = {2019},
  journal = {Meas. Sci. Technol.},
  volume = {30},
  pages = {085303},
  issn = {0957-0233, 1361-6501},
  doi = {10.1088/1361-6501/ab2587},
  url = {https://iopscience.iop.org/article/10.1088/1361-6501/ab2587},
  urldate = {2019-11-21},
  langid = {english},
  number = {8}
}

@article{champagne_turbulence_1967,
  title = {Turbulence Measurements with Inclined Hot-Wires {{Part}} 2. {{Hot-wire}} Response Equations},
  author = {Champagne, F. H. and Sleicher, C. A.},
  year = 1967,
  month = apr,
  journal = {J. Fluid Mech.},
  volume = {28},
  number = {1},
  pages = {177--182},
  copyright = {https://www.cambridge.org/core/terms},
  langid = {english}
}

@article{champagne_turbulence_1967a,
  title = {Turbulence Measurements with Inclined Hot-Wires {{Part}} 1. {{Heat}} Transfer Experiments with Inclined Hot-Wire},
  author = {Champagne, F. H. and Sleicher, C. A. and Wehrmann, O. H.},
  year = 1967,
  month = apr,
  journal = {J. Fluid Mech.},
  volume = {28},
  number = {1},
  pages = {153--175},
  langid = {english}
}

@book{hinze_turbulence_1975,
  title = {Turbulence},
  author = {Hinze, J. O.},
  year = 1975,
  publisher = {McGraw-Hill},
  googlebooks = {DfRQAAAAMAAJ},
  isbn = {978-0-07-029037-2},
  langid = {english}
}

@article{webster_note_1962,
  title = {A Note on the Sensitivity to Yaw of a Hot-Wire Anemometer},
  author = {Webster, C. a. G.},
  year = 1962,
  month = jun,
  journal = {J. Fluid Mech.},
  volume = {13},
  number = {2},
  pages = {307--312},
  langid = {english}
}

@article{xia_highfidelity_2025a,
    title = {High-fidelity simultaneous measurements of velocity and temperature fluctuations using a dual-wire sensor within thermal boundary layers},
    volume = {36},
    issn = {0957-0233},
    language = {en},
    number = {5},
    urldate = {2025-08-03},
    journal = {Meas. Sci. Technol.},
    author = {Xia, Y and Rowin, W Abu and Deshpande, R and Marusic, I and Hutchins, N},
    month = apr,
    year = {2025},
    pages = {055302},
}

@inproceedings{nurwahyu2025generating,
  title={Generating spanwise flow without penalties},
  author={Aliffrananda, Muhammad Hafiz Nurwahyu and Wong, Jeremy and Camobreco, Christopher J and Xia, Yu and Hutchins, Nicholas and Chung, Daniel},
  booktitle={Division of Fluid Dynamics Annual Meeting},
  year={2025},
  organization={APS}
}

@article{xia_evaluation_2024,
  title = {Evaluation of Spatial Resolution Effects in Rough Wall-Bounded Turbulence},
  author = {Xia, Y. and Chung, D. and Marusic, I. and Hutchins, N. and Abu Rowin, W.},
  year = 2024,
  month = sep,
  journal = {Exp. Fluids},
  volume = {65},
  number = {10},
  pages = {150}
}

@article{viotti_streamwise_2009,
  title = {Streamwise Oscillation of Spanwise Velocity at the Wall of a Channel for Turbulent Drag Reduction},
  author = {Viotti, Claudio and Quadrio, Maurizio and Luchini, Paolo},
  year = 2009,
  month = nov,
  journal = {Phys. Fluids},
  volume = {21},
  number = {11},
  eprint = {0911.0114},
  primaryclass = {physics},
  pages = {115109},
  archiveprefix = {arXiv},
  langid = {english}
}

@article{quadrio_drag_2011,
  title = {Drag Reduction in Turbulent Boundary Layers by In-Plane Wall Motion},
  author = {Quadrio, Maurizio},
  year = 2011,
  journal = {Philos. Trans. R. Soc.},
  volume = {369},
  number = {1940},
  eprint = {41061599},
  eprinttype = {jstor},
  pages = {1428--1442}
}

@article{ricco_review_2021a,
  title = {A Review of Turbulent Skin-Friction Drag Reduction by near-Wall Transverse Forcing},
  author = {Ricco, Pierre and Skote, Martin and Leschziner, Michael A.},
  year = 2021,
  month = may,
  journal = {Prog. Aerosp. Sci.},
  volume = {123},
  pages = {100713}
}

@article{leschziner_frictiondrag_2020a,
  title = {Friction-{{Drag Reduction}} by {{Transverse Wall Motion}} -- {{A Review}}},
  author = {Leschziner, Michael A.},
  year = 2020,
  month = oct,
  journal = {J. Mech.},
  volume = {36},
  number = {5},
  pages = {649--663},
  langid = {english}
}

@article{quadrio_streamwisetravelling_2009,
  title = {Streamwise-Travelling Waves of Spanwise Wall Velocity for Turbulent Drag Reduction},
  author = {Quadrio, Maurizio and Ricco, Pierre and Viotti, Claudio},
  year = 2009,
  month = may,
  journal = {J. Fluid Mech.},
  volume = {627},
  pages = {161--178},
  copyright = {https://www.cambridge.org/core/terms},
  langid = {english}
}

@article{peet_theoretical_2009,
    title = {Theoretical prediction of turbulent skin friction on geometrically complex surfaces},
    volume = {21},
    issn = {1070-6631},
    url = {https://doi.org/10.1063/1.3241993},
    doi = {10.1063/1.3241993},
    number = {10},
    urldate = {2026-03-15},
    journal = {Phys. Fluids},
    author = {Peet, Yulia and Sagaut, Pierre},
    month = oct,
    year = {2009},
    pages = {105105},
}

@inproceedings{kramer_wavy_2010,
    address = {Chicago, Illinois},
    title = {Wavy riblets for turbulent drag reduction},
    isbn = {978-1-62410-140-3},
    url = {https://arc.aiaa.org/doi/10.2514/6.2010-4583},
    doi = {10.2514/6.2010-4583},
    language = {en},
    urldate = {2024-11-26},
    booktitle = {5th {Flow} {Control} {Conference}},
    publisher = {AIAA},
    author = {Kramer, Felix and Grueneberger, Rene and Thiele, Frank and Wassen, Erik and Hage, Wolfram and Meyer, Robert},
    month = jun,
    year = {2010},
}

@article{sasamori_experimental_2014,
    title = {Experimental study on drag-reduction effect due to sinusoidal riblets in turbulent channel flow},
    volume = {55},
    issn = {1432-1114},
    url = {https://doi.org/10.1007/s00348-014-1828-z},
    doi = {10.1007/s00348-014-1828-z},
    language = {en},
    number = {10},
    urldate = {2026-03-15},
    journal = {Exp. Fluids},
    author = {Sasamori, M. and Mamori, H. and Iwamoto, K. and Murata, A.},
    month = oct,
    year = {2014},
    pages = {1828},
}

@article{sasamori_parametric_2017,
    title = {Parametric {Study} on a {Sinusoidal} {Riblet} for {Drag} {Reduction} by {Direct} {Numerical} {Simulation}},
    volume = {99},
    issn = {1573-1987},
    url = {https://doi.org/10.1007/s10494-017-9805-2},
    language = {en},
    number = {1},
    urldate = {2026-03-15},
    journal = {Flow Turbul. Combust.},
    author = {Sasamori, M. and Iihama, O. and Mamori, H. and Iwamoto, K. and Murata, A.},
    month = jul,
    year = {2017},
    pages = {47--69},
}

@article{cafiero_manipulation_2024,
    title = {Manipulation of a turbulent boundary layer using sinusoidal riblets},
    volume = {984},
    issn = {0022-1120, 1469-7645},
    url = {https://www.cambridge.org/core/product/identifier/S0022112024002568/type/journal_article},
    doi = {10.1017/jfm.2024.256},
    language = {en},
    urldate = {2025-03-05},
    journal = {J. Fluid Mech.},
    author = {Cafiero, Gioacchino and Amico, Enrico and Iuso, Gaetano},
    month = apr,
    year = {2024},
    pages = {A59},
}

@article{fujita_measurement_1968,
    title = {Measurement of {Reynolds} {Stress} by a {Single} {Rotated} {Hot} {Wire} {Anemometer}},
    volume = {39},
    issn = {0034-6748},
    url = {https://doi.org/10.1063/1.1683670},
    doi = {10.1063/1.1683670},
    number = {9},
    urldate = {2025-11-20},
    journal = {Rev. Sci. Instrum.},
    author = {Fujita, Hajime and Kovasznay, Leslie S. G.},
    month = sep,
    year = {1968},
    pages = {1351--1355},
}

@article{sirivat_measurement_1989,
    title = {Measurement and interpretation of space-time correlation functions and derivative statistics from a rotating hot wire in a grid turbulence},
    volume = {7},
    issn = {1432-1114},
    url = {https://doi.org/10.1007/BF00193416},
    doi = {10.1007/BF00193416},
    language = {en},
    number = {6},
    urldate = {2026-03-16},
    journal = {Exp. Fluids},
    author = {Sirivat, A.},
    month = jun,
    year = {1989},
    pages = {361--370},
}

@article{jelly_phaseaveraged_2017,
    title = {Phase-averaged flow statistics in compressors using a rotated hot-wire technique},
    volume = {58},
    issn = {1432-1114},
    url = {https://doi.org/10.1007/s00348-017-2326-x},
    doi = {10.1007/s00348-017-2326-x},
    language = {en},
    number = {5},
    urldate = {2025-11-20},
    journal = {Exp. Fluids},
    author = {Jelly, T. O. and Day, I. J. and di Mare, L.},
    month = apr,
    year = {2017},
    pages = {48},
}

@article{philip_spatial_2013,
  title = {Spatial Averaging of Streamwise and Spanwise Velocity Measurements in Wall-Bounded Turbulence Using {$\vee$}- and \texttimes -Probes},
  author = {Philip, Jimmy and Baidya, Rio and Hutchins, Nicholas and Monty, Jason P and Marusic, Ivan},
  year = 2013,
  month = nov,
  journal = {Meas. Sci. Technol.},
  volume = {24},
  number = {11},
  pages = {115302},
  langid = {english}
}

@article{orlu_comparison_2013,
  title = {Comparison of Experiments and Simulations for Zero Pressure Gradient Turbulent Boundary Layers at Moderate {{Reynolds}} Numbers},
  author = {{\"O}rl{\"u}, Ramis and Schlatter, Philipp},
  year = 2013,
  month = jun,
  journal = {Exp. Fluids},
  volume = {54},
  number = {6},
  langid = {english}
}

@article{aburowin_piv_2025,
  title = {{{PIV}} Measurements within Roughness Grooves with Enhanced Wall Reflection Suppression},
  author = {Abu Rowin, Wagih and Manovski, Peter and Chung, Daniel and Hutchins, Nicholas},
  year = 2025,
  month = feb,
  journal = {Exp. Fluids},
  volume = {66},
  number = {3},
  pages = {54},
  langid = {english}
  }

@article{ding_acceleration_2024a,
  title = {Acceleration Is the Key to Drag Reduction in Turbulent Flow},
  author = {Ding, Liuyang and Sabidussi, Lena F. and Holloway, Brian C. and Hultmark, Marcus and Smits, Alexander J.},
  year = 2024,
  month = oct,
  journal = {PNAS},
  volume = {121},
  number = {43},
  pages = {e2403968121}
}

@article{ramani_assessment_2024,
  title = {An Assessment of Effective Slope as a Parameter for Turbulent Drag Prediction over Multi-Scaled Roughness},
  author = {Ramani, A. and Schilt, L. and Nugroho, B. and Busse, A. and Jelly, T. O. and Monty, J. P. and Hutchins, N.},
  year = 2024,
  month = may,
  journal = {Exp. Fluids},
  volume = {65},
  number = {6},
  pages = {78},
  langid = {english}
}

\clearpage
\appendix

\section{Aerodynamic interference of a vertical hot-wire configuration}
\label{app:vertical}

While deploying a vertical hot-wire configuration (stem parallel to the wall-normal $z$-axis) simplifies the rotational mechanism by eliminating the need for a Z-Shaped offset bracket, it introduces significant aerodynamic interference. In this orientation, the prongs are positioned perpendicular to the primary flow direction rather than aligned with it, creating a local blockage that artificially accelerates the flow over the sensor. 

\begin{figure}[!h]
    \centering
    \includegraphics{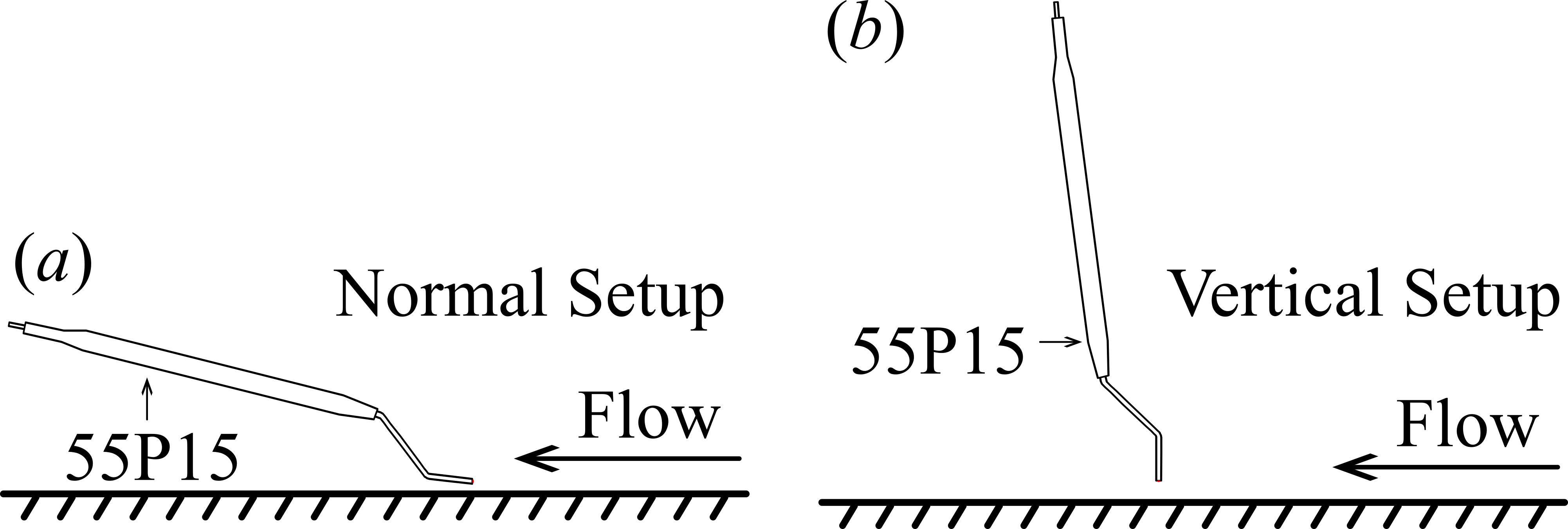}
    \caption{Schematics of the hot-wire probe deployment orientation: ($a$) normal setup and ($b$) vertical setup.}
    \label{fig:verticalHotwire}
\end{figure}

\begin{figure}[!h]
    \centering
    \includegraphics{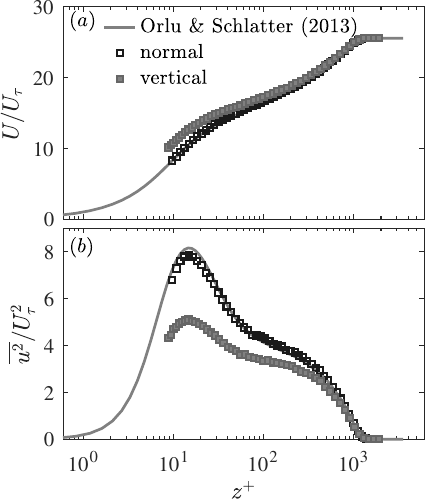}
    \caption{Profiles of mean ($a$) and variance ($b$) streamwise velocity comparing the normal and vertical hot-wire configurations over a smooth wall. The solid line denotes DNS data from \citet{orlu_comparison_2013} at a matched Reynolds number of $Re_\tau \approx 1150$.}
    \label{fig:verticalNormalhotwireResults}
\end{figure}

To quantify this effect, streamwise velocity measurements over a smooth wall were compared between a standard normal setup and a vertical setup (figure \ref{fig:verticalHotwire}$a$ and $b$, respectively).  Figure \ref{fig:verticalNormalhotwireResults}($a$) illustrates that while the normal configuration matches the DNS data of \citet{orlu_comparison_2013} at $Re_\tau \approx 1150$, the vertical configuration exhibtis a distinct overestimation of the mean velocity ($U/U_\tau$) in the near-wall region. This artificial local acceleration is accompanied by a substantial attenuation of the streamwise velocity variance, as shown in figure \ref{fig:verticalNormalhotwireResults}($b$). Consequently, the vertical deployment is unsuitable for accurate boundary layer profiling, necessitating the Z-bracket assembly described in Section \ref{sec:rotatingProbe}.

\section{Influence of spanwise drift during probe rotation}
\label{app:Driftrotating}
\begin{figure}[!h]
    \centering
 \includegraphics{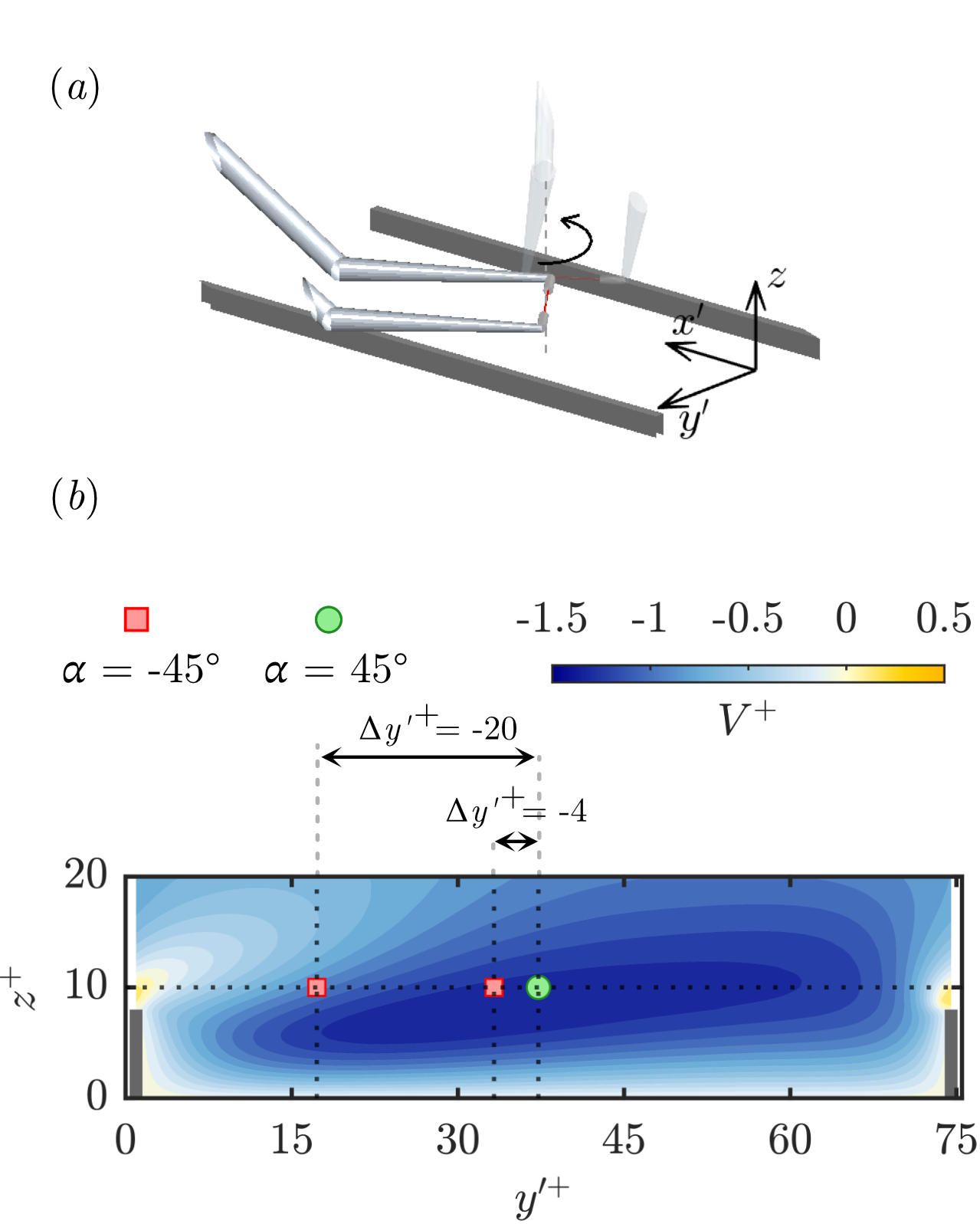}
    \caption{($a$) Schematic showing probe's position at $\alpha = 45^o$ (dark) and $\alpha = -45^o$ (light) over a wide-yawed riblet surface, with $x'$ parallel and $y'$ orthogonal to the riblet blades. ($b$) Cross section of wide-yawed riblet surface in the $y'$--$z$ plane  showing the spanwise position of the probe's centre at $\alpha = 45^{\circ} $ (circle) and $\alpha = -45^{\circ} $ (squares). Two squares represent two cases corresponding to spanwise drift $\Delta y'^+$ of -4 and -20. Contours in the background represent the distribution of the mean spanwise velocity ($V^+$) obtained from the DNS database.}
    \label{fig:hwmisalign}
\end{figure}

The implementation of the Z-shaped bracket introduces the potential for eccentric rotation if the hot-wire filament centre is not perfectly collinear with the rotational axis of the stepper motor shaft. During rotation from $\alpha = 45^\circ$ to $-45^\circ$ (a full $90^\circ$ sweep), any mechanical misalignment manifests as spatial drift. Without delving into the kinematics of the spatial drift, here we simply highlight the sensitivity of spanwise-drifting hot-wire by simulating the measurement with the DNS database.

Figure \ref{fig:hwmisalign}($a$) shows the rotating probe over a wide-yawed riblet surface in the riblet-aligned local ($x'$, $y'$) coordinate system. The centre of the filament is aligned at the local spanwise centre when  $\alpha = 45^\circ$. For the purpose of illustration, we assume that the hot-wire probe linearly drifts along $-ve$ $y'$ direction as it rotates to $\alpha = -45^\circ$. Figure \ref{fig:hwmisalign}($b$) shows the position of the probe's centre in the $y'-z$ cross-section at $z^+ = 10$ when $\alpha = 45^{\circ}$ (green circle) and $\alpha = -45^{\circ}$ (red squares) for two representative cases, $\Delta y'^+ = -4$ and $\Delta y'^+ = -20$. The selection of negative values of $\Delta y'^+$ is to highlight the influence of stronger non-uniformity of the mean spanwise velocity ($V^+$) as evidenced by the contours plotted in the cross-section using the DNS database. The riblet blades induce local secondary flows, causing the mean spanwise velocity to vary along the $y'$-direction in the near-wall region. The sensitivity of hot-wire probe to spatial drift stems from this spanwise non-uniformity of the flow bounded by the riblet blades. While the spanwise flow is relatively uniform and strong in the central region between the blades, its magnitude reduces near the blade walls, even exhibiting an opposing spanwise flow ($V^+ > 0$) in the leeward side of the blade. Consequently, if the hot-wire drifts too close to the blades during rotation, it measures these localised flow variations rather than the bulk spanwise velocity, which subsequently distorts the $\overline{u^2_e}$ profiles. 

\begin{figure}[!h]
    \centering
    \includegraphics{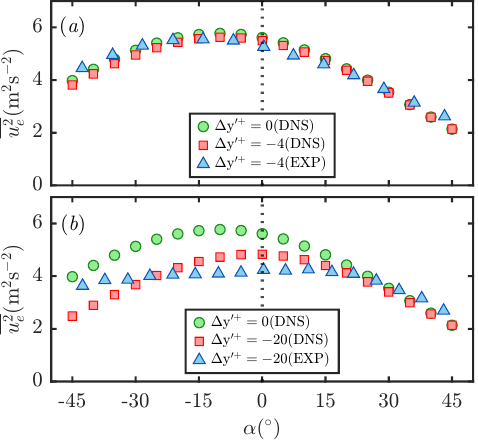}
    \caption{Mean-squared effective velocity $\overline{u_e^2}$ as a function of yaw angle subjected to spanwise drift of ($a$) $\Delta y'^+ = -4$ and ($b$) $\Delta y^+ = - 20$ (red squares), together with a baseline case $\Delta y'^+ = 0$ (green circles), simulated at the same wall-normal location ($z^+ = 10$). Also shown are the experimental data (blue triangles) for corresponding spanwise drifts.}
    \label{fig:driftsensitivity}
\end{figure}

For example, consider a hot-wire with an active sensor length of $l^+ \approx 11$ positioned in the spanwise centre of the yawed riblet ($\varphi = 22.3^{\circ}$) for $\alpha = 45^{\circ}$. Figure \ref{fig:driftsensitivity} plots the mean-squared effective velocity $\overline{u_e^2}$ as a function of yaw angle subjected to spanwise drift of ($a$) $\Delta y'^+ = -20$ and ($b$) $\Delta y^+ = - 4$ (red squares), together with a baseline case $\Delta y'^+ = 0$ (green circles), simulated using DNS database at the wall-normal location, $z^+ \approx 10$. Also, test experiments were carried where the probe was allowed to drift spanwise by $\Delta y^+ \approx - 4$ and $\Delta y^+ \approx - 20$ at the same wall normal location. The corresponding experimental data are presented as well using blue triangles. Figure \ref{fig:driftsensitivity}($a$) shows that if the misalignment error is properly contained limiting the spanwise drift to $\Delta y'^+ \approx -4$, the distortion in measurements is minimal, yielding a clean effective velocity profile that accurately captures the flow angle. However, if the misalignment error is inadequately controlled with spanwise drift reaching to $\Delta y'^+ \approx -20$, then figure \ref{fig:driftsensitivity}($b$) shows that the measurements at different angular positions will be distorted compared to the baseline case ($\Delta y'^+ = 0$). This will compromise the least-squares fit of Eq. \ref{eq:ChampagneEalpha}, and we fail to accurately identify the peak yaw position and measure the true spanwise velocity. Also, the experimental data does not collapse well with the DNS simulated data for $\Delta y'^+ \approx -20$. This is because the probe drifts to the region of unsteady three dimensional flows near the blade compromising its directional sensitivity. Strict concentricity is therefore essential during rotation for reliable spanwise velocity measurements.

\section{Derivation of the spatial Stokes layer}\label{app:SSL}

The classical SSL solutions \citep{viotti_streamwise_2009} are based on periodic boundary conditions formulated with a finite wavelength $\lambda_x$. \RVA{The straight-to-yawed riblet transition,
by contrast, acts as a localised step change in the wall condition, from no-slip to a
spanwise wall translation.} To capture the spatial development of this forcing directly in
the physical domain, the downstream evolution of the mean spanwise velocity $V(x,z)$ is governed by the linearised equation:
\begin{equation}
  U_{z,\mathrm{w}}\,z\,\frac{\partial V}{\partial x}=\nu\frac{\partial^2 V}{\partial z^2},
  \label{eq:linearAssumption}
\end{equation}
where $U_{z,\mathrm{w}} = U^2_\tau/\nu$ is the mean streamwise velocity gradient at the wall.
\RVA{
Equation~\eqref{eq:linearAssumption} holds while the forcing remains within the region where
the streamwise profile is approximately linear; its accuracy therefore degrades once the
penetration depth grows beyond the viscous sublayer.
}
A similarity transformation reduces \eqref{eq:linearAssumption} to an ordinary differential
equation (ODE),
\begin{equation}
  V(x,z) = V_\mathrm{equ}\,F(\eta), \quad \text{where} \quad \eta = \frac{z}{\delta_m(x)}.
  \label{eq:similarSolution}
\end{equation}
Here, $\delta_m(x)$ is a similarity length scale. Substituting Eq. \eqref{eq:similarSolution}
into Eq. \eqref{eq:linearAssumption} and applying the chain rule gives
\begin{equation}
  \frac{F''(\eta)}{\eta^2 F'(\eta)}
  = -\frac{U_{z,\mathrm{w}}}{\nu}\,\delta_m^2\,\frac{\mathrm{d}\delta_m}{\mathrm{d}x}
  = -C_r.
  \label{eq:similarSolution2}
\end{equation}
\RVA{
Since both sides depend on the independent variables ($\eta$ and $x$), they must equal a separation constant $-C_r$. In similarity frameworks, the choice of this constant is arbitrary. We set
$C_r = 3$ so that the integration below yields an $\exp(-\eta^3)$ kernel.} Eq.
\eqref{eq:similarSolution2} then decouples into a growth-rate equation and the similarity ODE,
\begin{flalign}
  & \delta_m(x) = \left(\frac{9\nu x}{U_{z,\mathrm{w}}}\right)^{1/3}, \label{eq:mathGrowthRate}\\
  & F''(\eta) + C_r\,\eta^2 F'(\eta) = 0. & \label{eq:similarityODE}
\end{flalign}
For a semi-infinite step-up impulse simulating the straight-to-yawed transition, the
boundary conditions are $F(0)=1$ at the wall and $F(\infty)=0$ in the far field. Integrating Eq.
\eqref{eq:similarityODE} subject to these conditions yields the closed-form similarity solution
\begin{equation}
  V_\mathrm{step}(x,z) = V_\mathrm{equ}\left(1 - \frac{1}{\Gamma(4/3)}
  \int_0^{\eta}\exp(-r^3)\,\mathrm{d}r\right),
  \label{eq:SSLsolution1}
\end{equation}
where $\Gamma(\cdot)$ is the gamma function and $r$ is the variable of integration. Defining the penetration depth $\delta_p$ as the
height at which the spanwise velocity has decayed by $99\%$
($V_\mathrm{step}/V_\mathrm{equ}=0.01$) yields an upper limit $\eta_{99}\approx1.404$, and hence
\begin{equation}
  \delta_p \approx 1.404\left(\frac{9\nu^2 x}{U^2_\tau}\right)^{1/3}.
  \label{eq:SSLthickness2}
\end{equation}
Non-dimensionalising by viscous wall units, $\delta_m^+ = (9x^+)^{1/3}$, so the similarity
variable becomes $\eta = z^+/(9x^+)^{1/3}$, yielding
\begin{align}
  & V_{\mathrm{step}}^+(x^+,z^+) = V_{\mathrm{equ}}^+\left(1 - \frac{1}{\Gamma(4/3)}
    \int_{0}^{\eta}\exp(-r^3)\,\mathrm{d}r\right), \label{eq:VstepApp}\\
  & \delta^+_p(x^+) \approx 2.92\,(x^+)^{1/3}. & \label{eq:dpApp}
\end{align}
Eq. \eqref{eq:VstepApp} is the response to the onset of forcing at a single edge, where the yawed riblets begin to drive the spanwise flow at $V_\mathrm{equ}^+$. Where the yawed riblets end, the forcing is removed and the wall returns to rest, which is the same step taken with the opposite sign. As Eq.~\eqref{eq:linearAssumption} is linear, a finite yawed patch follows by superposing the onset and the removal of the forcing.

Only the yawed riblets drive the spanwise flow; straight riblets and the smooth wall do not. Each tile is then a single yawed patch, beginning at $x_s^+$ and ending at $x_e^+$, and the same description applies whether the patch sits on straight riblets or on the smooth wall. With the Heaviside step function ($\mathcal{H}=1$ for a positive argument and $0$ otherwise),
\begin{align}
  V^+(x^+, z^+) &= \mathcal{H}(x^+ - x_s^+)\,V_{\mathrm{step}}^+(x^+ - x_s^+, z^+) \nonumber\\
  &\quad - \mathcal{H}(x^+ - x_e^+)\,V_{\mathrm{step}}^+(x^+ - x_e^+, z^+).
  \label{eq:patchApp}
\end{align}
Downstream of the trailing edge ($x^+ > x_e^+$), where the forcing has been removed, both steps are active and Eq.~\eqref{eq:patchApp} reduces to
\begin{equation}
  V^+(x^+, z^+) = \frac{V_\mathrm{equ}^+}{\Gamma(4/3)}\int_{\eta_s}^{\eta_e}\exp(-r^3)\,\mathrm{d}r,
  \label{eq:patchSimpApp}
\end{equation}
with
\begin{align}
  \eta_s = \frac{z^+}{[9(x^+ - x_s^+)]^{1/3}}, \qquad
  \eta_e = \frac{z^+}{[9(x^+ - x_e^+)]^{1/3}}.
  \label{eq:etaSE}
\end{align}
Here $\eta_e$ sets the internal layer $z_\mathrm{inner}^+\approx2.92(x^+-x_e^+)^{1/3}$, which grows from the wall once the forcing is removed, and $\eta_s$ sets the outer reach of the upstream forcing, $z_\mathrm{outer}^+\approx2.92(x^+-x_s^+)^{1/3}$. Below $z_\mathrm{inner}^+$ the
fluid has returned to rest; between $z_\mathrm{inner}^+$ and $z_\mathrm{outer}^+$ it retains the spanwise momentum from the patch as a detached shear layer.

\end{document}